# Data-Driven Multiscale Design of Cellular Composites with Multiclass Microstructures for Natural Frequency Maximization

Liwei Wang [a,b], Anton van Beek [b], Daicong Da [b], Yu-Chin Chan [b],
Ping Zhu [a,*], Wei Chen [b,*]

a. The State Key Laboratory of Mechanical System and Vibration, Shanghai Key Laboratory of Digital Manufacture for Thin-Walled Structures, School of Mechanical Engineering, Shanghai Jiao Tong University, Shanghai, P.R. China

b. Department of Mechanical Engineering, Northwestern University, 2145 Sheridan RD Tech B224, Evanston, IL 60201, USA

*Corresponding authors
Email address: pzhu@sjtu.edu.cn
Email address: weichen@northwestern.edu

**ABSTRACT**

For natural frequency optimization of engineering structures, cellular composites have been shown to possess an edge over solid. However, existing multiscale design methods for cellular composites are either computationally exhaustive or confined to a single class of microstructures. In this paper, we propose a data-driven topology optimization (TO) approach to enable the multiscale design of cellular structures with various choices of microstructure classes. The key component is a newly proposed latent-variable Gaussian process (LVGP) model through which different classes of microstructures are mapped into a low-dimensional continuous latent space. It provides an interpretable distance metric between classes and captures their effects on the homogenized stiffness tensors. By introducing latent vectors as design variables, a differentiable transition of stiffness matrix between classes can be easily achieved with an analytical gradient. After integrating LVGP with the density-based TO, an efficient data-driven cellular composite optimization process is developed to enable concurrent exploration of microstructure concepts and the associated volume fractions for natural frequency optimization. Examples reveal that the proposed cellular designs with multiclass microstructures achieve higher natural frequencies than both single-scale and single-class designs. This framework can be easily extended to other multi-scale TO problems, such as thermal compliance and dynamic response optimization.

## 1. INTRODUCTION

Cellular composites have shown advantages over their solid counterparts in a wide variety of multifunctional engineering structure designs. Amongst these applications, natural frequency optimization is of great importance to avoid destructive vibrational resonance. Various topology optimization (TO) methods have been proposed to design the macroscale structure for frequency optimization [1], such as the solid isotropic material with penalization (SIMP) [2-6], the evolutionary structural optimization (ESO) [7-10] and the level set method (LSM) [11, 12]. While these works focus on macroscale design, Liu et al. assumed a full structure composed of periodic unit cells and performed design optimization at the microscale [13]. To fully exploit the hierarchical nature of cellular composites in natural frequency optimization, the body of literature on multiscale TO for cellular composites has grown to consider both macro- and microscale designs.

Based on the assumed microstructure distributions, state-of-the-art multiscale design methods for cellular structure design can be categorized into three types. The first type of method assumes a periodic distribution of microstructures within the full structure and then performs optimization in two scales concurrently [14-17]. While this periodicity assumption can greatly reduce the number of microscale designs for higher efficiency, it also imposes a strict limitation on the design space that fails to meet spatially varying property requirements and thus leads to sub-optimal designs. To enlarge the design space, the second type of method associates each element or elemental integration point with an independent microscale design in the multiscale optimization [18, 19]. However, a dramatic increase in the number of design variables together with the nested optimization will result in an unaffordable computational cost. Also, it is difficult to ensure good connections between neighboring microstructures. As an attempt to compromise between computation cost and design freedom, the third type of method assumes that the full structure is composed of several subregions of periodic designs. The division of subregions is achieved based on some heuristic criteria [20-23] or multi-material optimization methods [24, 25]. With a small number of subregions, it is easy to handle the compatibility between microstructures by adding non-design kinematic connectors [20, 24] or connectivity constraints [26] to the design process. However, as the number of subregions increases, its advantage of efficiency will quickly diminish.

With the growth of data resources and computational power, data-driven TO has become a promising method to achieve higher efficiency in larger design spaces. Most existing data-driven multiscale TO methods are based on the framework of variable-density cellular structure design [27-31]. In this framework, the full structure is assumed to be composed of the same form (class) of microstructures with varying density (volume fraction) of the solid materials. A surrogate model can be trained to approximate the relation between the volume fraction and the homogenized stiffness tensor of the microstructure, such as neural network [29], scaling law model [27, 28], and diffuse approximation [30]. Using this surrogate model to replace the material law, elemental volume fraction can be used as the only design variable in the multiscale design without the need to perform on-the-fly homogenization and nested optimization. After obtaining the optimal volume fraction distribution, the full structure can be readily assembled by microstructures with spatially varying porosity. Although this framework is simple and efficient, using a predefined class of microstructures sacrifices the generality of the design. To address this issue, Wang et al. [32] developed a parameterized interpolation scheme between two specially selected classes of microstructures to obtain a richer microstructure library. Zhang et al. [33] proposed to interpolate a set of key prototype microstructures to form a family of variable-density cellular composites during the optimization process. Nevertheless, the interpolation, homogenization, and on-the-fly training of the surrogate model compromises the efficiency of the optimization. In addition, it confines regions with the same volume fraction to have the same microstructure. This neglects the fact that these regions might be under different deformation in general design cases and would impose different property requirements. Liu et al. considered multiple classes of microstructures in a library based on the multi-material design framework [34]. However, it introduces high-dimensional design variables for each elemental microstructure design that greatly lowers the efficiency. Liu et al. enabled data-driven design with multiple series of lattice unit cells based on heuristic selection principles, which is more applicable for static compliance minimization instead of the dynamic optimization problem [35].

The aim of this study is thus to develop a data-driven multiscale TO approach that can consider multiple classes of microstructures for better dynamic performance without much loss of efficiency. In this study, as shown in **Fig. 1**, each element or subdomain of the full structure is described by both quantitative (e.g., volume fraction) and qualitative variables (e.g., the class of microstructure). By considering the class of microstructure as a qualitative variable, the dimension of the design variables is greatly reduced while retaining large design freedom. However, different discrete classes do not have a well-defined distance metric, resulting in a challenging mixed-variable problem for both surrogate modeling and gradient-based TO. To account for this multiclass microstructure behavior, we employ our recently proposed latent variable Gaussian process (LVGP) [36] to transform qualitative classes into continuous latent variables based on their effects in the homogenized properties represented by the stiffness tensor. These latent variables are combined with the quantitative volume fraction as inputs in creating the surrogate model of unit cell geometry-property relations.

In meeting a broader range of property targets, we include microstructures with normal-shear coupling terms in the stiffness tensor the are weakly correlated with the other entries. As a result, a conventional separable kernel structure in LVGP falls short in describing this weak correlation and leads to poor prediction accuracy. To mitigate this issue, we propose to use the sum of separable (SoS) kernels for a more flexible covariance structure in accomodating normal-shear coupling terms. This newly developed LVGP model is integrated with the density-based TO method to enable efficient data-driven design for natural frequency maximization that can simultaneously explore multiple classes of microstructures and associated volume fractions. The benefits of our method are demonstrated by the achieved higher natural frequencies than both single-scale and single-class designs.

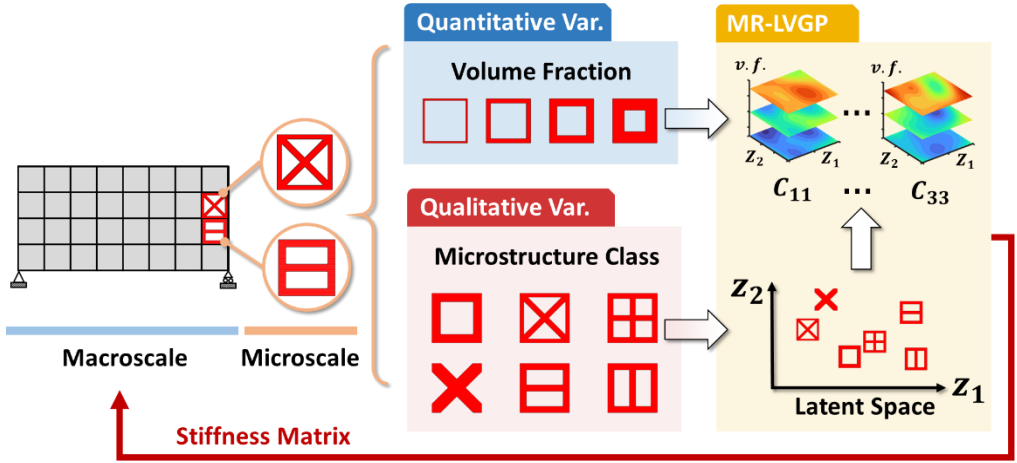

**Fig. 1** Illustration of the proposed approach

The remaining paper is organized as follows. In Section 2, we describe how to generate a precomputed library with multiclass microstructures. The latent variable Gaussian process with the sum of separable kernels (LVGP-SoS) is then proposed to enable surrogate modeling for the pre-computed multiclass library. Integrating the LVGP-SoS model with the TO framework is outlined in Section 3 to maximize the natural frequencies of the full structure. To demonstrate the effectiveness of the proposed method, numerical design examples are provided in Section 4. Finally, we draw the conclusion and discuss future works in Section 5.

## 2. MULTICLASS MICROSTRUCTURE LIBRARY CONSTRUCTION AND SURROGATE MODELING

In this section, we first describe the procedure of constructing a library with prescribed classes of microstructures. A novel latent variable Gaussian process with the sum of separable kernels is then introduced to enable accurate surrogate modeling of the homogenized stiffness tensor given mixed-variable inputs. By using this model, discrete classes of microstructures are mapped into a continuous latent space to facilitate the multiscale TO design in the next section.

### 2.1. Multiclass microstructure generation and property calculation

It is advisable to include multiple classes of microstructure to cover a wide range of homogenized properties, i.e. stiffness tensor, to accommodate general design requirements. In this study, 10 classes of parameterized lattice models are used to generate microstructures included in the library, as shown in **Fig. 2 (a)**. We choose these classes because most of the optimized multiscale TO designs in existing literature contains one of these microstructure configurations [16, 17, 33, 34]. In fact, it can be noted that these lattice models include various combinations of rods in different directions, rendering diverse symmetry types and directional characteristics of the homogenized stiffness, as shown in **Fig. 2 (b)**. Though these microstructures have distinct geometrical patterns, they are preselected to connect with each other, ensuring manufacturability of the assembled structure. Herein, we associate each lattice model with a geometric parameter $a$ to describe the detailed geometry of microstructures. Given a target volume fraction of solid material, the analytical expression for the geometric parameter $a$ can be readily obtained for each class to determine the corresponding microstructures.

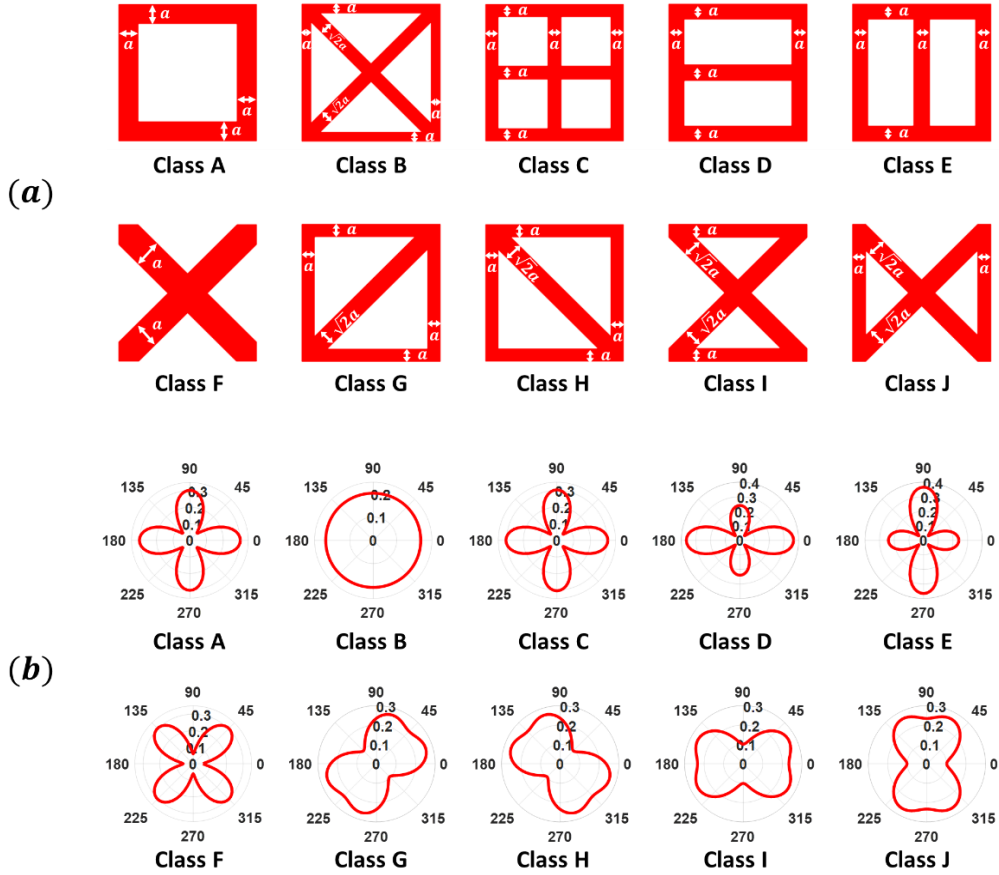

**Fig. 2** Illustration of the selected classes, (a) geometry definition, (b) elasticity surfaces of each class by rotating the stiffness matrix to obtain $C_{11}$ values in different directions.

To facilitate the later property calculation, the microstructure is discretized into a $200 \times 200$ pixel matrix, with zero and one to represent void and solid, respectively. We sample volume fraction values for each class and compute the corresponding geometric parameter $a$ to generate microstructures, establishing a library with 10 classes and 795 microstructures. Energy-based homogenization [37] is then used to calculate their effective stiffness tensors. Herein, we set the Poisson's ratio to be 0.3 and Young's modulus to be 201GPa for the base material. To facilitate surrogate modeling, independent entries of the stiffness tensor, i.e., $C_{11}, C_{12}, C_{13}, C_{22}, C_{23}$, and $C_{33}$, are normalized with respect to Young's modulus and plotted in **Fig. 3** with the volume fraction as the horizontal axis.

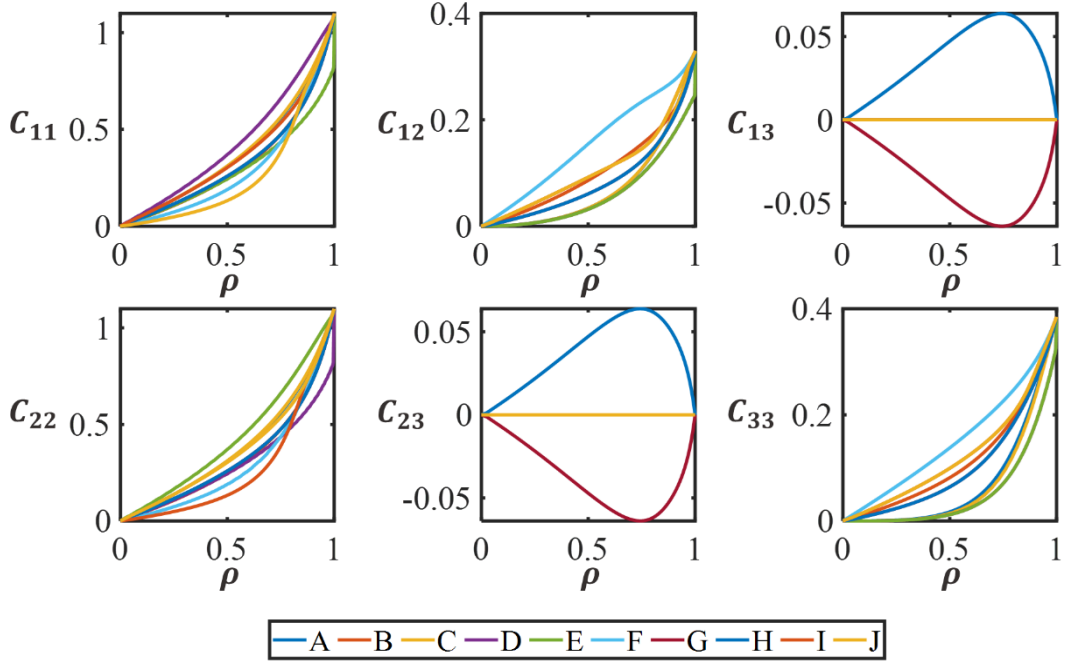

**Fig. 3** The relations between independent entries of stiffness tensor and the volume fraction values for different classes.

From **Fig.3**, we can see that entries of the stiffness tensor generally change smoothly with little nonlinearity as the volume fraction increases. Therefore, this mid-size database is enough to train the surrogate model with satisfactory accuracy. It should be pointed out that entries $C_{13}$ and $C_{23}$ describe the ***coupling effect*** between the shear and normal directions while the other entries only relate to strain and stress in the same direction. This normal-shear coupling phenomenon only exists for microstructures without orthotropic symmetry, which is the case for classes G and H. As a result, $C_{13}$ and $C_{23}$ change non-monotonically for classes G and H as the volume fraction increases but remain zero for other classes. In contrast, other entries have a similar monotonic curve for all classes of microstructures.

**2.2. Latent Variable Gaussian Process with Sum of Separable Kernels**

While including multiple microstructure classes in the constructed library renders a larger property space than in existing works, it also introduces surrogate modeling challenges. Specifically, the class of microstructure is a qualitative variable that serves as a nomenclature without any well-defined distance metric. As a result, it is difficult to define correlation structures for surrogate modeling and compute the gradient for optimization. Moreover, as discussed in Section 2.1, the stiffness tensor contains two normal-shear coupling terms that behave differently from the other entries, imposing another difficulty for multi-response surrogate modeling.

To address these challenges, we propose an LVGP-SoS model in this study based on two simple but powerful concepts: latent variable and sum of separable kernels. As shown in **Fig 4**, for a mixed-variable physical model, there are always some underlying quantitative physical variables $\boldsymbol{V} = \{v_1, v_2, \dots, v_n\} \in \mathrm{R}^n$ to explain or represent the effects of a qualitative factor $t$ on the response. While these underlying variables can be extremely high-dimensional, we could assume a low-dimensional and continuous latent space $\boldsymbol{z}$ to capture their joint effects. In other words, we could map qualitative factors into a continuous low-dimensional latent space based on their effects on the response, which is the key idea for our recently proposed latent variable Gaussian process (LVGP) [38].

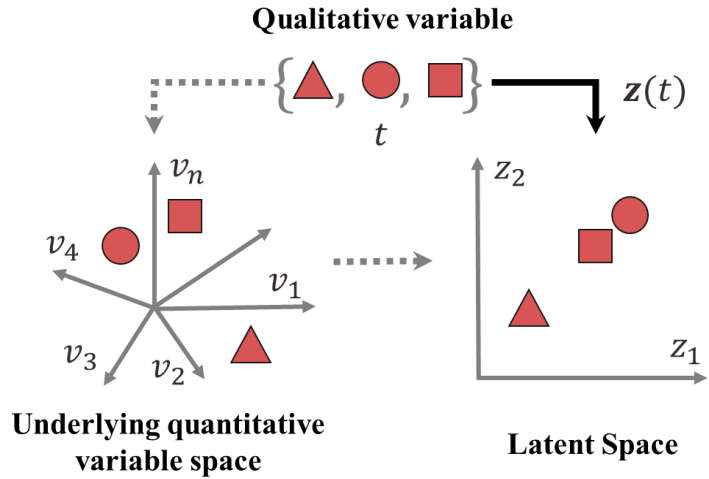

**Fig. 4** Illustration of the underlying variables and latent space of qualitative variables.

Specifically, consider a computer simulation model $y(w)$ with output $y = [y_1, y_2, \ldots, y_{N_{op}}]^T \in R^{N_{op}}$ and input $w = [x^T, t^T]^T$ containing both quantitative variables $x = [x_1, x_2, \ldots, x_p]^T \in R^p$ and qualitative variables $t = [t_1, t_2, \ldots, t_q]^T$, with the $j^{th}$ qualitative factor $t_j \in \{1, 2 \ldots, l_j\}$. Herein, $l_j \in N^+$ is the total number of levels for the $j^{th}$ qualitative factor $t_j$. Based on the idea of latent representation, we assume that each qualitative variable $t_j$ is mapped to a $g$-dimensional latent vector $z_j(t_j) = [z_{j,1}(t_j), \ldots, z_{j,g}(t_j)]^T \in R^g$. Denote the transformed input vector as $s = [x^T, z(t)^T]^T \in R^{p+q \times g}$, where $z(t) = [z_1(t_1)^T, \ldots, z_q(t_q)^T]^T$. We can then assume the prior model for response $y_i$ to be

$$Y_i(s) = \gamma_i + G_i(s), \tag{1}$$

where $\gamma_i$ is a constant mean function and $G_i(\cdot)$ is a zero-mean Gaussian process. In our previous work [39, 40], a separable structure is adopted for the covariance to enable multi-response modeling:

$$cov\left(G_i(s), G_j(s')\right) = \Sigma_{ij} \cdot r(s, s'), \tag{2}$$

where $\Sigma_{ij}$ is the covariance term between the pair of responses $y_i$ and $y_j$, and $r(\cdot, \cdot)$ is a spatial correlation function defined between two transformed input vectors $s$ and $s'$. By using this separable structure, the same correlation function is applied to every output of the model. While this practice works well for the stiffness prediction of orthotropic microstructures in our previous work [39, 40], it will be problematic for the ***non-orthotropic microstructure*** classes considered in this study due to the special normal-shear coupling terms. To solve this, we propose to use an independent correlation function for each entry of the stiffness tensor to increase the modeling flexibility in accommodating coupling terms, which can be viewed as a special case of the sum of separable kernels [41, 42]. With independent correlation functions, the covariance for $G_i(\cdot)$ is changed to

$$cov(G_i(s), G_i(s')) = c_i(s, s') = \sigma_i^2 r_i(s, s'), \tag{3}$$

where $\sigma_i^2$ is the prior variance, $c_i(\cdot,\cdot)$ is the covariance function, and $r_i(\cdot,\cdot)$ is the independent correlation function for the response $y_i$. Herein, we adopt the most commonly used Gaussian correlation function:

$$r_i(s, s') = \exp\left\{-(x - x')^T \Phi_i (x - x') - (z(t) - z(t'))^T \Phi_z (z(t) - z(t'))\right\}, \quad (4)$$

where $\Phi_i = diag(\phi_i)$ and $\phi_i = [\phi_{i,1}, \phi_{i,2}, \ldots, \phi_{i,p}]^T$ are scaling parameters of the quantitative variable to be estimated for response $y_i$, $\Phi_z = diag(\phi_z)$ and $\phi_z = [\phi_{z,1}, \phi_{z,2}, \ldots, \phi_{z,q}]^T$ are undecided scaling parameters of the latent variables shared among all the responses. By using this structure, each response can have independent scaling parameters for the quantitative variables to better describe its respective behavior. Meanwhile, different responses still have a unified latent space since the latent variables and their scaling parameters are shared. As a result, the latent space remains low-dimensional and highly interpretable, which is desirable for later optimization.

Hyperparameters to be estimated in this newly proposed model are $z, \Phi_z, \{\Phi_i\}, \{\gamma_i\}$ and $\{\sigma_i^2\}$. For a size-$n$ training dataset with observed inputs $W = [w^{(1)}, w^{(2)}, \ldots, w^{(n)}]^T$ and response data $D = [y^{(1)}, y^{(2)}, \ldots, y^{(n)}]^T = [d_1, d_2, \ldots, d_{N_{op}}]$, the corresponding log-likelihood function of the model is

$$L_{ln}(Z, \Phi_z, \{\Phi_i\}, \{\gamma_i\}, \{\sigma_i^2\}) = \sum_{i=1}^{N_{op}} -\frac{n}{2}\ln(\sigma_i^2) - \frac{1}{2}\ln(|R(Z, \Phi_z, \Phi_i)|)$$
$$- \frac{1}{2\sigma_i^2}(d_i - \mathbf{1}\gamma_i)^T R(Z, \Phi_z, \Phi_i)^{-1}(d_i - \mathbf{1}\gamma_i), \quad (5)$$

where $\ln(\cdot)$ is the natural logarithm, $\mathbf{1}$ is an $n \times 1$ vector of ones, $R$ is the $n \times n$ correlation matrix with $R_{ij} = r(s^{(i)}, s^{(j)})$ for $i, j = 1, \ldots, n$, and $Z = \bigcup_{i=1}^{q}\{z_i(1), \ldots, z_i(l_i)\}$ is the set of mapped latent variable values for all the levels of the qualitative variables. With this log-likelihood function, the maximum likelihood estimates (MLEs) for $\gamma_i$ and $\sigma_i^2$ can be obtained as

$$\hat{\gamma}_i = \frac{\mathbf{1}^T R(Z, \boldsymbol{\Phi}_z, \boldsymbol{\Phi}_i)^{-1} \boldsymbol{d}_i}{\mathbf{1}^T R(Z, \boldsymbol{\Phi}_z, \boldsymbol{\Phi}_i)^{-1} \mathbf{1}}, \tag{6}$$

$$\hat{\sigma}_i^2 = \frac{1}{n}(\boldsymbol{d}_i - \mathbf{1}\hat{\gamma}_i)^T R(Z, \boldsymbol{\Phi}_z, \boldsymbol{\Phi}_i)^{-1}(\boldsymbol{d}_i - \mathbf{1}\hat{\gamma}_i). \tag{7}$$

After substituting (6) and (7) into (5), the MLEs $\hat{Z}$, $\hat{\boldsymbol{\Phi}}_z$ and $\hat{\boldsymbol{\Phi}}_i$ can be obtained by minimizing the negative log-likelihood function (ignore constant terms):

$$[\hat{Z}, \hat{\boldsymbol{\Phi}}_z, \hat{\boldsymbol{\Phi}}_i] = \operatorname*{argmin}_{Z, \boldsymbol{\Phi}_z, \boldsymbol{\Phi}_i} \sum_{i=1}^{N_{op}} n\ln(\hat{\sigma}_i^2) + \ln(|R(Z, \boldsymbol{\Phi}_z, \boldsymbol{\Phi}_i)|). \tag{8}$$

The prediction for the response $y_i$ can then be obtained at a given input $\boldsymbol{w}^* = [\boldsymbol{x}^{*T}, \boldsymbol{t}^{*T}]^T$ through

$$\hat{y}_i(\boldsymbol{w}^*) = \hat{y}_i(\boldsymbol{s}^*) = \hat{\gamma}_i + \boldsymbol{r}_i^T R(\hat{Z}, \hat{\boldsymbol{\Phi}}_z, \hat{\boldsymbol{\Phi}}_i)^{-1}(\boldsymbol{d}_i - \mathbf{1}\hat{\gamma}_i), \tag{9}$$

where $\boldsymbol{r}_i = [r_i(\boldsymbol{s}^*, \boldsymbol{s}^{(1)}), r_i(\boldsymbol{s}^*, \boldsymbol{s}^{(2)}), \ldots, r_i(\boldsymbol{s}^*, \boldsymbol{s}^{(n)})]^T$, and $\boldsymbol{s}^* = [\boldsymbol{x}^{*T}, \boldsymbol{z}(\boldsymbol{t}^*)^T]^T$ is the transformed input. Due to the Gaussian correlation function adopted in the model, the prediction function is guaranteed to be differentiable. Therefore, the analytical expressions of partial derivatives $\frac{\partial \hat{y}_i}{\partial \boldsymbol{x}}$ and $\frac{\partial \hat{y}_i}{\partial \boldsymbol{z}}$ can be easily obtained from (9). For more detailed illustrations and implementation of the ordinal Gaussian process and LVGP modeling, readers are referred to [38, 43, 44].

**2.3. Surrogate Modeling for Multiclass Microstructure Library**

In this section, the proposed LVGP-SoS model is applied in surrogate modeling for the stiffness matrices of microstructures, which takes the volume fraction $\rho$ as a quantitative input and the class of microstructures $t$ as a qualitative input. Considering the symmetry of the stiffness matrix, we only need to use the independent entries $y = [C_{11}, C_{12}, C_{13}, C_{22}, C_{23}, C_{33}]^T$ as the output vector for the model. Since we will use the latent vectors as design variables in the later optimization process, a low-dimensional latent space is preferred to avoid computational challenges caused by the high dimensionality and sparsity. Based on our previous works [38], a 2D latent space is generally sufficient for most physical models, which is also adopted in this study. As a comparative study, we train an LVGP-SoS model as well as a multi-response LVGP model with the separable kernel structure (denoted as MR-LVGP) on the constructed library divided into training (80%) and test (20%) sets. The training results are shown in **Table 1**.

**Table 1** Fitting results of the LVGP-SoS and MR-LVGP models

|  | LVGP-SoS (Proposed) | | | MR-LVGP | | |
| --- | --- | --- | --- | --- | --- | --- |
|  | MSE | RMSE | $r^2$ | MSE | RMSE | $r^2$ |
| $C_{11}$ | $5.41 \times 10^{-5}$ | $7.35 \times 10^{-3}$ | 0.9994 | $1.58 \times 10^{-3}$ | 0.0398 | 0.9815 |
| $C_{12}$ | $9.57 \times 10^{-6}$ | $3.09 \times 10^{-3}$ | 0.9988 | $2.64 \times 10^{-4}$ | 0.0163 | 0.9665 |
| $C_{13}$ | $5.48 \times 10^{-10}$ | $2.34 \times 10^{-5}$ | 0.9999 | $2.16 \times 10^{-5}$ | 0.0046 | 0.9373 |
| $C_{22}$ | $5.30 \times 10^{-5}$ | $7.28 \times 10^{-3}$ | 0.9994 | $1.68 \times 10^{-3}$ | 0.0410 | 0.9803 |
| $C_{23}$ | $5.48 \times 10^{-10}$ | $2.34 \times 10^{-5}$ | 0.9999 | $2.16 \times 10^{-5}$ | 0.0046 | 0.9373 |
| $C_{33}$ | $4.74 \times 10^{-6}$ | $2.18 \times 10^{-3}$ | 0.9996 | $3.69 \times 10^{-4}$ | 0.0192 | 0.9668 |

From the results, it can be noted that the MR-LVGP model with a conventional separable kernel has a lower prediction accuracy for the two normal-shear coupling terms $C_{13}$ and $C_{23}$, as expected. In contrast, the proposed LVGP-SoS model achieves a much higher prediction accuracy for all entries despite the distinct homogenized behaviors of different classes. Specifically, the coefficients of determinations ($r^2$) are larger than 0.99 while the mean square errors (MSE) and root mean square errors (RMSE) are smaller than $10^{-4}$ and $10^{-2}$, respectively. This demonstrates the advantage of introducing the sum of separable kernels in the LVGP model. Moreover, the LVGP-SoS model also provides a highly interpretable latent space for different classes of microstructures, as shown in **Fig. 5 (a)** and **(b)**.

This latent space induces a natural distance metric between classes that encodes the correlation between their responses. The larger the distance in the latent space between two classes of microstructures, the weaker the correlation between them in terms of their responses, and vice versa. For example, classes A and C are closed to each other in the latent space since their homogenized stiffness tensors are nearly the same. Classes with crossed rods at the center (classes B, F, I, and J) have a relatively small mutual distance in the latent space. This matches well with the observation that these classes have a better resistance for the shear deformation compared with the remaining classes. Classes D, E, G, and H include extra rods to reinforce the stiffness in a certain direction. Correspondingly, they surround other classes in the latent space with a large radius. As another demonstration of the interpretability of the latent space, we fix the volume fraction to be 0.5 and then obtain the elasticity modulus surfaces for points uniformly sampled in the latent space, as shown in **Fig. 6**.

It is noted that the modulus surface has a continuous and smooth change in the latent space. Therefore, the gradient information of the stiffness tensor can be readily obtained through LVGP-SoS in the latent space, which is otherwise unavailable for the original qualitative class. This interpretable latent space with rich physical information is highly desirable for the data-driven multiscale TO, which will be demonstrated in the remaining parts of this paper.

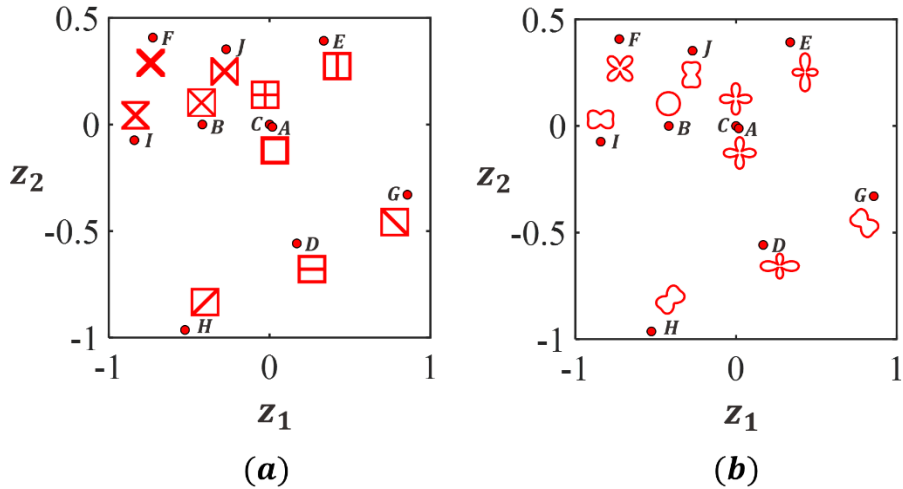

**Fig. 5** Latent spaces for the fitted LVGP-SoS model, (a) latent space marked by geometries, (b) latent space marked by elasticity modulus surfaces.

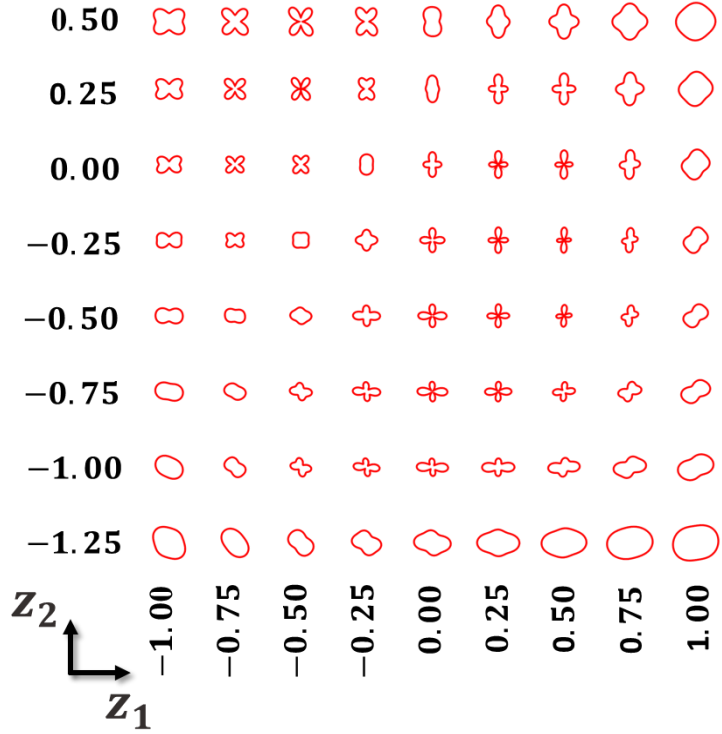

**Fig. 6** Predicted elasticity surfaces for sampling points in the latent space

## 3. DATA-DRIVEN MULTISCALE TOPOLOGY OPTIMIZATION

In this section, we illustrate how to integrate the proposed LVGP-SoS model with density-based TO to enable efficient multiscale TO with multiclass microstructures. Based on this integrated framework, the sensitivity analysis of the multiscale system is introduced for maximizing the natural frequencies. We then outline the numerical implementation for this multiscale method.

### 3.1. TO Integrated with LVGP-SoS

As illustrated in the last section, the proposed LVGP-SoS serves as an accurate surrogate model for the stiffness tensor and renders a continuous latent embedding for classes of microstructures, inducing a well-defined distance metric to capture their collective effects on the response. By varying the latent variables, a smooth and continuous transformation of stiffness tensor between classes can be achieved.

Therefore, we can readily integrate the LVGP-SoS model into the multiscale TO by including the latent vector and quantitative variables (e.g., volume fraction) as design variables and employing LVGP-SoS to efficiently obtain the homogenized stiffness tensor and its gradient.

Specifically, the trained Co-LVGP model is denoted as $Y(\rho, z(t))$, where $\rho$ is the volume fraction of the microstructure, $t$ is the class of microstructures, $z = [z_1, z_2]^T$ is a 2D latent vector corresponding to $t$, and $Y$ is a vector of independent entries of the stiffness matrix. Given a combination of $\rho$, $z_1$ and $z_2$, the geometry of the microstructure is determined, together with the homogenized stiffness matrix predicted by the LVGP-SoS model $k(\rho, z_1, z_2) = k(Y(\rho, z_1, z_2))$. To ensure the feasibility of the optimized structure, we need to introduce two restrictions for the design space.

The first restriction is imposed on the volume fraction $\rho$ to avoid thin rods in the microstructures. This is achieved by filtering the volume fraction field of the structure during the optimization through the projection scheme given as [45]

$$\tilde{\rho} = \rho \cdot \widetilde{H}(\rho, \eta, \beta), \qquad (10)$$

where $\widetilde{H}$ is an approximated Heaviside function [46],

$$\widetilde{H}(\rho, \eta, \beta) = \frac{tanh(\beta\eta) + tanh(\beta(\rho - \eta))}{tanh(\beta\eta) + tanh(\beta(1 - \eta))}. \qquad (11)$$

$\beta$ and $\eta$ are parameters used to control the sharpness and threshold of the projection, respectively. By adding this filtering technique, those low $\rho$ values will be automatically filtered out during the optimization. We denote the corresponding mass matrix after the filtering to be $m(\tilde{\rho}(\rho)) = m(\rho).$

The second restriction is imposed on latent vectors. Although the LVGP-SoS model can provide a stiffness matrix for any given latent vector, only those vectors mapped from the prescribed classes have corresponding lattice models. To drive the latent vectors to these existing classes, the stiffness matrix is multiplied by an extra penalization term $f(z_1, z_2)$:

$$\widetilde{\boldsymbol{k}}(\rho, z_1, z_2) = f(z_1, z_2)\boldsymbol{k}(\widetilde{\rho}(\rho), z_1, z_2), \tag{12}$$

$$f = \exp\left\{-1/\alpha \cdot \min_t\left(\|\boldsymbol{z} - \boldsymbol{z}(t)\|_2^2\right)\right\}$$
$$= \max_t\left\{\exp\left(-1/\alpha \cdot \|\boldsymbol{z} - \boldsymbol{z}(t)\|_2^2\right)\right\}, \tag{13}$$

where $f: R^2 \to (0,1]$ is the penalty function and $\alpha$ is a decay parameter to control the level of penalization. The larger the decay parameter, the greater the penalization will be. To facilitate sensitivity analysis and optimization, the penalty function is further approximated by a smoothed expression as

$$f = 1/\lambda \cdot ln\left\{\sum_t \exp\left(\lambda \cdot \exp(-1/\alpha \cdot \|\boldsymbol{z} - \boldsymbol{z}(t)\|_2^2)\right)\right\}, \tag{14}$$

where $\lambda$ is a large constant. Herein, we set $\lambda$ to be 500 and $\alpha$ to be the diagonal length of the minimum bounding box for the existing latent vectors. Parameter $\alpha$ can also be gradually decreased during the optimization to accelerate convergence. In this study, we decrease $\lambda$ by 5% in each iteration until it reaches 0.01.

With these two extra restrictions, we discretize the design region into a mesh of four-node quadrilateral elements. Each element is assigned with three design variables (one density $\rho^{(e)}$ and two latent variables $z_1^{(e)}, z_2^{(e)}$) to determine the corresponding microstructure, the elemental stiffness matrix $\widetilde{\boldsymbol{k}}_e\left(\rho^{(e)}, z_1^{(e)}, z_2^{(e)}\right)$ and the elemental mass matrix $\boldsymbol{m}_e(\rho^{(e)})$. The natural frequencies of the designed structure are obtained by solving the following general eigenvalue problem

$$(\boldsymbol{K} - \omega_i^2 \boldsymbol{M})\boldsymbol{u}_i = \boldsymbol{0}, \tag{15}$$

where $\omega_i$ is the *i*th natural frequency (eigenvalue), $\boldsymbol{u}_i$ is the eigenmode (eigenvector) corresponding to $\omega_i$, $\boldsymbol{K}$ and $\boldsymbol{M}$ are the global stiffness matrix and mass matrix, respectively. We can rewrite this formulation into the form of the Rayleigh quotient

$$\omega_i^2 = \frac{\boldsymbol{u}_i^T \boldsymbol{K} \boldsymbol{u}_i}{\boldsymbol{u}_i^T \boldsymbol{M} \boldsymbol{u}_i}. \tag{16}$$

In this study, we consider the multiscale TO problem to maximize the natural frequency under a given constraint of the overall base material volume fraction, which can be formulated as

$$\max_{\boldsymbol{\rho},\mathbf{z}_1,\mathbf{z}_2} \omega_i(\boldsymbol{\rho},\mathbf{z}_1,\mathbf{z}_2)$$

$$s.t.\ (\mathbf{K} - \omega_i^2\mathbf{M})\boldsymbol{u}_i = \mathbf{0},$$

$$V_* - \frac{1}{N}\sum_{e=1}^{N}\tilde{\rho}^{(e)} = 0, \quad (17)$$

$$z_i^- \leq z_i^{(e)} \leq z_i^+, i = 1,2,$$

$$0 < \rho_{min} \leq \rho^{(e)} \leq 1,$$

where $N$ is the total number of elements, $\tilde{\rho}^{(e)}$ is the filtered elemental volume fraction value, $V_*$ is the target solid material volume fraction of the full structure, $z_i^-$ ($z_i^+$) is the lower (upper) bound for the $i^{th}$ latent variable and $\rho_{min}$ is a small value $(10^{-6})$ to avoid singularity.

### 3.2. Sensitivity Analysis

By taking the derivative of (16), the sensitivity of the objective function can be obtained as

$$\frac{\partial \omega_i}{\partial \rho^{(e)}} = \frac{1}{2\omega_i \boldsymbol{u}_i^T \mathbf{M}\boldsymbol{u}_i}\left[2\frac{\partial \boldsymbol{u}_i^T}{\partial \rho^{(e)}}(\mathbf{K}-\omega_i^2\mathbf{M})\boldsymbol{u}_i + \boldsymbol{u}_i^T\left(\frac{\partial \mathbf{K}}{\partial \rho^{(e)}} - \omega_i^2\frac{\partial \mathbf{M}}{\partial \rho^{(e)}}\right)\boldsymbol{u}_i\right],$$

$$\frac{\partial \omega_i}{\partial z_j^{(e)}} = \frac{1}{2\omega_i \boldsymbol{u}_i^T \mathbf{M}\boldsymbol{u}_i}\left[2\frac{\partial \boldsymbol{u}_i^T}{\partial z_j^{(e)}}(\mathbf{K}-\omega_i^2\mathbf{M})\boldsymbol{u}_i + \boldsymbol{u}_i^T\left(\frac{\partial \mathbf{K}}{\partial z_j^{(e)}}\right)\boldsymbol{u}_i\right], j=1,2. \quad (18)$$

Substituting (15) into the above equations and assuming that the eigenmode $\boldsymbol{u}_i$ is normalized with respect to $\mathbf{M}$, the sensitivity expressions can be simplified as

$$\frac{\partial \omega_i}{\partial \rho^{(e)}} = \frac{1}{2\omega_i}\left[\boldsymbol{u}_i^T\left(\frac{\partial \mathbf{K}}{\partial \rho^{(e)}} - \omega_i^2\frac{\partial \mathbf{M}}{\partial \rho^{(e)}}\right)\boldsymbol{u}_i\right],$$

$$\frac{\partial \omega_i}{\partial z_j^{(e)}} = \frac{1}{2\omega_i}\left[\boldsymbol{u}_i^T\left(\frac{\partial \mathbf{K}}{\partial z_j^{(e)}}\right)\boldsymbol{u}_i\right], j=1,2, \quad (19)$$

where the derivatives of the global mass matrix $\mathbf{M}$ and stiffness matrix $\mathbf{K}$ can be calculated as

$$\frac{\partial \mathbf{M}}{\partial \rho^{(e)}} = \frac{\partial \boldsymbol{m}_e}{\partial \tilde{\rho}^{(e)}}\frac{\partial \tilde{\rho}^{(e)}}{\partial \rho^{(e)}}$$

$$\frac{\partial \mathbf{K}}{\partial \rho^{(e)}} = f\left(z_1^{(e)}, z_2^{(e)}\right) \sum_i \frac{\partial \boldsymbol{k}_e}{\partial Y_i}\frac{\partial Y_i}{\partial \tilde{\rho}^{(e)}}\frac{\partial \tilde{\rho}^{(e)}}{\partial \rho^{(e)}}, \quad (20)$$

$$\frac{\partial \mathbf{K}}{\partial z_j^{(e)}} = f\left(z_1^{(e)}, z_2^{(e)}\right) \sum_i \frac{\partial \boldsymbol{k}_e}{\partial Y_i}\frac{\partial Y_i}{\partial z_j^{(e)}} + \frac{\partial f}{\partial z_j^{(e)}} \boldsymbol{k}_e, j = 1,2,$$

where $\frac{\partial \boldsymbol{m}_e}{\partial \tilde{\rho}^{(e)}}$ and $\frac{\partial \boldsymbol{k}_e}{\partial Y_i}$ are constant matrix obtained from the shape function of the four-node quadrilateral elements, $\frac{\partial Y_i}{\partial \tilde{\rho}^{(e)}}$ and $\frac{\partial Y_i}{\partial z_j^{(e)}}$ can be calculated by direct differentiation of (9), $\frac{\partial f}{\partial z_j^{(e)}}$ and $\frac{\partial \tilde{\rho}^{(e)}}{\partial \rho^{(e)}}$ can be formulated as

$$\frac{\partial f}{\partial z_j^{(e)}} = -\frac{2}{\alpha A} \cdot \sum_t \exp\left(\lambda \cdot \exp\left(-\frac{\|\boldsymbol{z} - \boldsymbol{z}(t)\|_2^2}{\alpha}\right) - \frac{\|\boldsymbol{z} - \boldsymbol{z}(t)\|_2^2}{\alpha}\right)\left(z_j^{(e)} - z_j(t)\right), (21)$$

$$\frac{\partial \tilde{\rho}^{(e)}}{\partial \rho^{(e)}} = \widetilde{H}(\rho^{(e)}, \eta, \beta) + \rho^{(e)} \cdot \frac{\beta\left(1 - tanh\left(\beta(\rho^{(e)} - \eta)\right)\right)^2}{tanh(\beta\eta) + tanh(\beta(1 - \eta))}. \quad (22)$$

It should be noted that we only consider simple eigenfrequency (algebraic multiplicity equals one) in this study, which guarantees the validity of (18). For the case with the multiple eigenfrequency, the derivative of the eigenfrequency cannot be obtained through (18) due to the lack of differentiability of the subspace spanned by the eigenvectors [9]. The mathematical perturbation analysis method proposed by Seyranian et al. [47] can be implemented to handle the multiple eigenfrequency.

### 3.3. Numerical Implementation of Multiscale TO

Based on the optimization problem definition and sensitivities analysis, we propose a sequential three-stage data-driven method to enable multiscale optimization with multiclass microstructures, as illustrated in **Fig. 7**.

In the first stage, $\rho^{(e)}, z_1^{(e)}, z_2^{(e)}$ of each element are taken as design variables. Therefore, the class of microstructure and its volume fraction are simultaneously optimized. In this stage, we only use the penalized stiffness matrix illustrated in (12) but do not implement the low-volume-fraction filtering technique given in (10). This is achieved by simply setting $\tilde{\rho}^{(e)} = \rho^{(e)}$ and $\frac{\partial \tilde{\rho}^{(e)}}{\partial \rho^{(e)}} = 1$. We employ the method of moving asymptotes (MMA) [48] to solve the optimization problem (16) iteratively based on the sensitivity obtained from (19). The sensitivity filtering technique is implemented to avoid numerical instability and possible defects. Similar to the issue of intermediate density in the SIMP method, the penalization on the stiffness matrix will encourage latent vectors to converge to those discrete points representing existing classes but not guarantee an exact convergence. The possible resultant "intermediate classes" do not correspond to any prescribed lattice models and can cause an inaccurate result due to the penalization. This issue is addressed in the next stage.

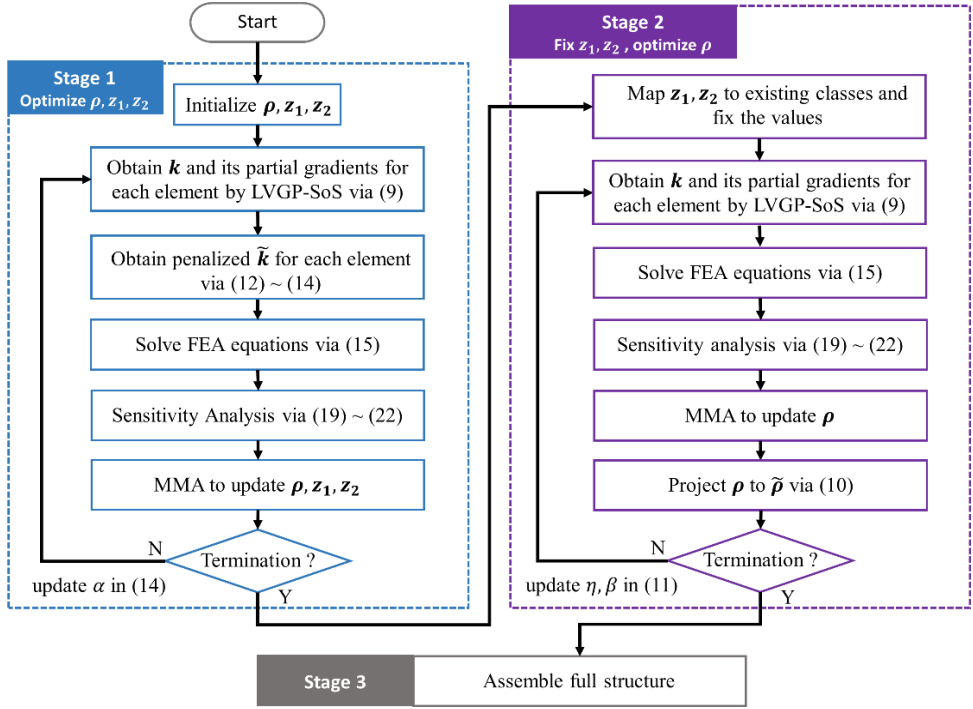

**Fig. 7** Flowchart of the three-stage method.

In the second stage, the $z_1$ and $z_2$ vectors of the optimized structure in the first stage will be projected to the nearest classes in the latent space. With the resultant structure as the initial solution, the TO process will be performed once again with the elemental volume fraction as the only design variable. Different from the first stage, the volume fraction projection scheme given in (10) will be used in this second-stage optimization. Specifically, volume fraction values lower than 0.06 will be filtered to avoid microstructures with very thin rods. To obtain a better result and enhance the numerical stability, we will gradually increase $\eta$ and $\beta$ to 0.06 and 400 during the optimization as suggested in [45], respectively. The continuation scheme for these two projection parameters is shown in **Fig. 8**, updating every 40 optimization iterations. From **Fig. 8**, it can be noted that the projection function first has its threshold gradually shifted to 0.06 and then increases the sharpness of the Heaviside function. This renders a gradual contraction of the design space that can help the optimization to converge to a better solution and improve the numerical stability. In the last stage, microstructures correspond to the optimal design variables obtained from the second stage will be used to assemble the full structure.

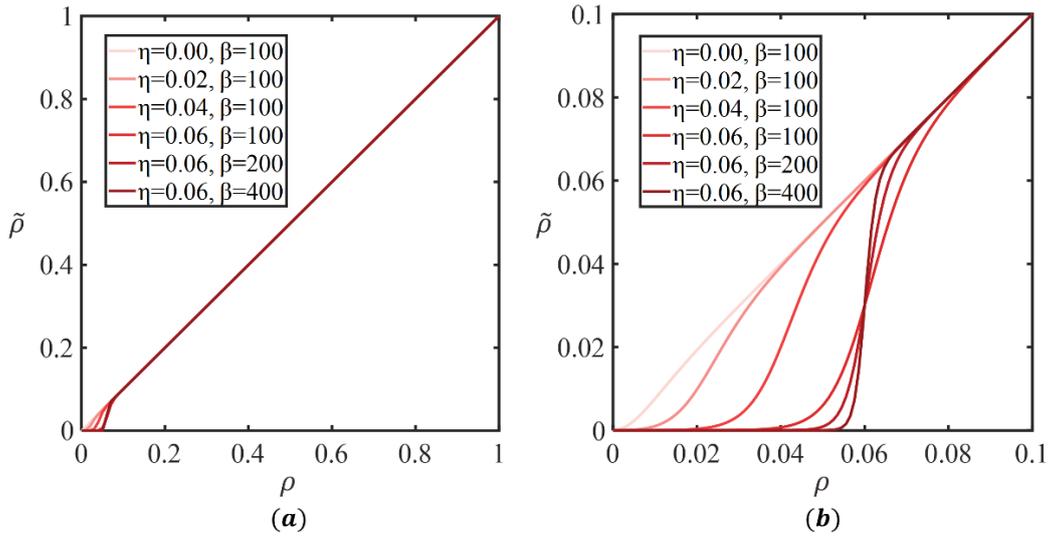

**Fig. 8** Projection scheme of the volume fraction for different combinations of $\eta$ and $\beta$ following the order of the continuation, (b) is an enlarged plot of (a).

## 4. DESIGN CASE STUDIES

To demonstrate the effectiveness and characteristics of the proposed data-driven design method, three design examples are presented in this section. All the design examples focus on maximizing the fundamental (first) natural frequency, which is a critical quantity for engineering structures to avoid destructive resonance. However, the proposed method can also be applied to the maximization of other frequencies. In the following examples, the base material is the same as the one used in Section 2 for the construction of the microstructure library. It has a Poisson's ratio of $v = 0.3$, the elastic modulus of $E = 201$GPa, and a density of $\rho_0 = 2700 \ kg/m^3$. As an initial guess for the first stage, each element is set to be filled by the class A microstructure with the elemental volume fraction equal to the target global volume fraction $V_*$. The first stage optimization will terminate when the change in design variables (normalized) is less than $0.01$ in two consecutive iterations or the number of iterations reaches $100$. The termination criteria for the second-stage optimization is that the change in elemental volume fraction is less than $0.01$ or the number of iterations reaches $250$.

**4.1. Clamped Beam with a Single Concentrated Mass**

As the first example, we aim to maximize the first natural frequency of a clamped beam as shown in **Fig. 9**. The beam is clamped at its left and right ends with a concentrated lump mass $m_c = 2000 \ kg$ imposed at the center. The $1.4 \ m \times 0.2 \ m$ design domain is equally discretized into $210 \times 30$ quadrilateral elements. The volume fraction of the full structure is set to be 40%.

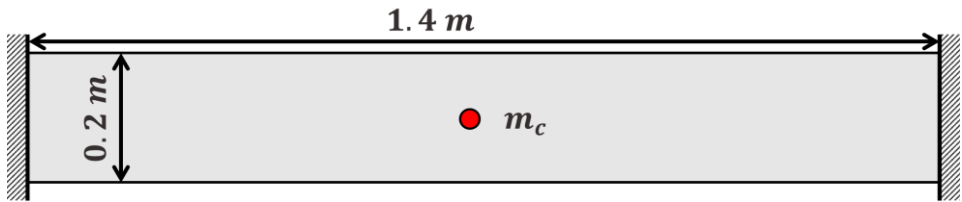

**Fig. 9** Problem setting illustration of the first example.

The optimization history and the design result are given in **Fig. 10**. From the optimization history in **Fig. 10 (a)**, it is observed that the algebraic multiplicity of the first natural frequency remains one during the whole optimization process. Also, there is no mode switch between the first and second frequencies. Therefore, the sensitivity analysis used in this study is valid. It can be noted that the first natural frequency has a large increase at the beginning of the first stage and then slightly decreases. This is due to the way we impose the penalty on the stiffness matrix. In the beginning, the parameter $\alpha$ is relatively large, corresponding to a weak penalization on the stiffness matrix for latent variables other than the 10 latent vectors mapped from existing classes. Therefore, the algorithm can freely explore the whole latent space, resulting in a large increase in the objective function. In the later part of the first stage, the penalization reaches a higher level as the parameter $\alpha$ decreases. The exploration in the latent space is gradually confined to the neighboring regions of existing classes. As a result, we observe a small decrease in the first frequency. After that, the objective function quickly becomes stable in the first stage and only has a small change in the second stage. This indicates that our penalization technique has successfully driven the latent variables to those points corresponding to predefined classes. In terms of efficiency, our data-driven approach replaces numerous microscale designs and the costly on-the-fly homogenization process with an inexpensive and accurate mixed-variable Gaussian process model. In addition, by using the latent variables, we only need to include two extra design variables for each element to represent multiple classes of microstructures. As a result, each iteration of the optimization process can have an efficiency comparable to the macro-scale TO. Moreover, we can observe that the objective function and volume fraction quickly enter a stable state within 60~80 iterations in the first stage, which is efficient considering the multiclass microstructure behavior. In contrast, if a standard TO is adopted to achieve the same resolution of the multiscale design in this study, it will require around 1300 times as many design variables and a continuation scheme of density projection with much more iterations [45].

The optimal design in **Fig. 10 (b)** has a frame with high-volume-fraction microstructures, which is in line with the prevailing design result of a double-clamped beam in the literature [7, 11]. Regions surrounded by the frame are filled by microstructures with relatively low volume fractions. From **Fig. 10 (c)**, we can observe that these low-volume-fraction microstructures in different classes can be allocated in a way that can better resist the local distortion patterns induced by the first eigenmode shown in **Fig. 10 (d)**. This validates the effectiveness of the proposed method.

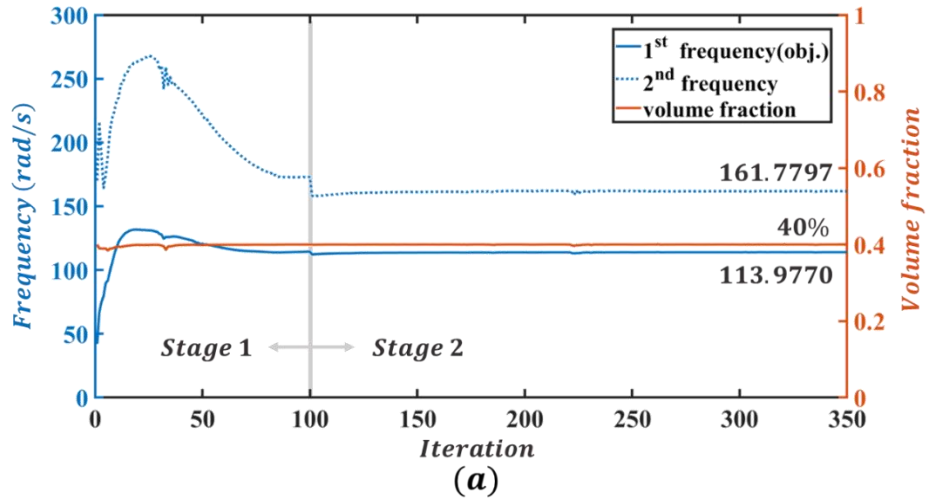

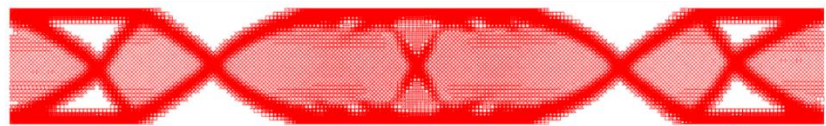

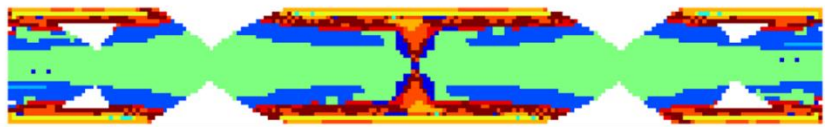

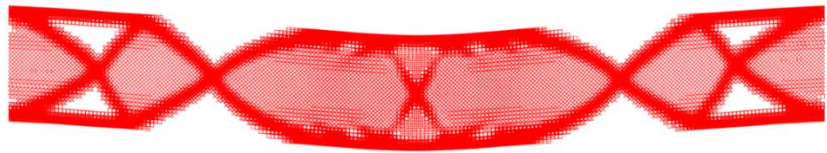

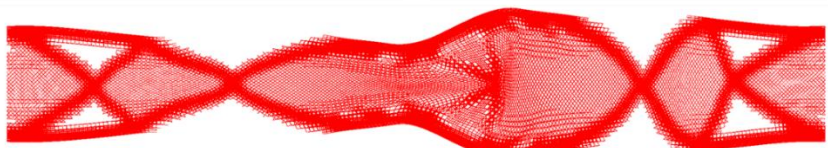

**Fig. 10** Design results of the first example with the proposed method, (a) is the optimization history, (b) and (c) are the assembled full structure and its distribution of different classes, respectively, (d) shows the first (up) and the second (bottom) eigenmodes.

To demonstrate the advantages of this multi-class design method, we compare the optimal result with the single-class and single-scale designs. For all single-class designs in this study, we assume the full structure is composed of the same class of microstructure and only optimize the volume fraction distribution. As a result, there will be 10 different single-class designs corresponding to the 10 classes included in the library. We select the best structures among these candidates to represent the optimal single-class design. As for single-scale designs in this study, the SIMP method [46] with Heaviside projection is employed, setting the penalty exponent to be 3. The structure designed by SIMP is then mapped to a 0-1 design with the volume fraction unchanged. The design results for this example are shown in **Fig. 11**.

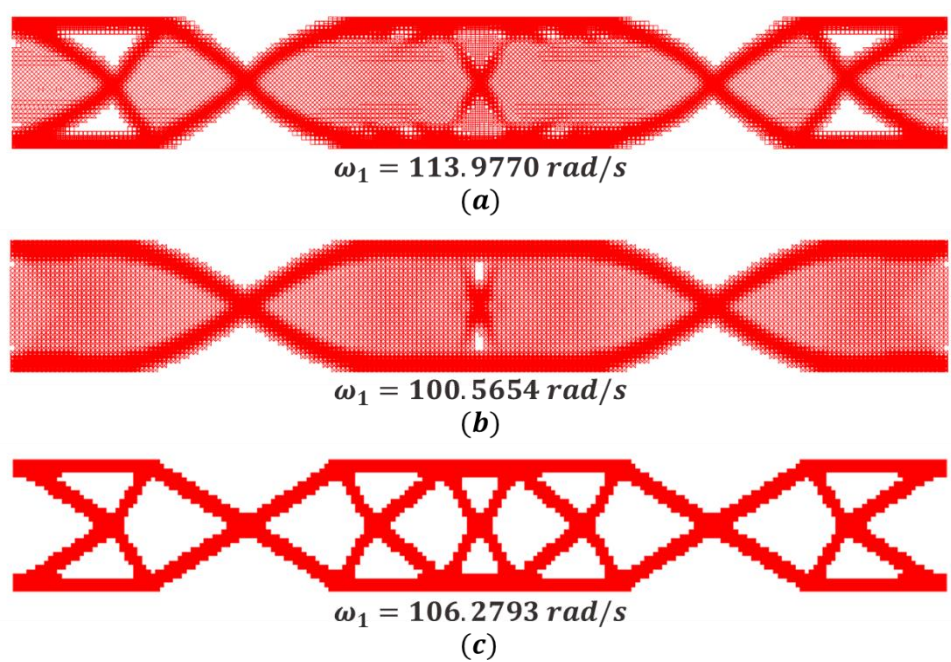

**Fig. 11** Comparison of structures obtained by different methods, (a) multiscale design with multi-class microstructures, (b) multiscale design with single-class microstructures, (d) single-scale design via SIMP.

From the figure, we conclude that, although three designs have similar solid frame structures, multi-class optimization outperforms both single-class and single-scale design. Specifically, the first frequency of the multi-class design is improved by 13.34% and 7.24% over single-class and single-scale designs, respectively. This is different from the case for static compliance design where multiscale cellular design offers no extra benefit [22, 49]. The result also shows that the fundamental eigenfrequency of the single-class design is much smaller than both multi-class and single-scale design. This underlines the importance of combining different classes to accommodate various local property requirements. In contrast, using a single predefined microstructure design concept for the whole structure, which is very common in existing literature [17, 28, 30], will be suboptimal for general design cases.

**4.2. Pinned Beam with a Single Concentrated Mass**

In this example, the design region is a beam pinned at both ends with a concentrated lump mass $m_c = 2000\ kg$ imposed at the middle of the bottom, as illustrated in **Fig. 12**. The $1.0\ m \times 0.3\ m$ design domain is equally discretized into $100 \times 30$ quadrilateral elements. The volume fraction of the full structure is set to be 45%. Similarly, we compare the performance of single-scale, single-class, and multi-class designs. In this example, we further explore the influence of the number of classes included in the optimization and therefore perform the multi-class optimization with different numbers of classes, i.e., six, eight, and 10 classes.

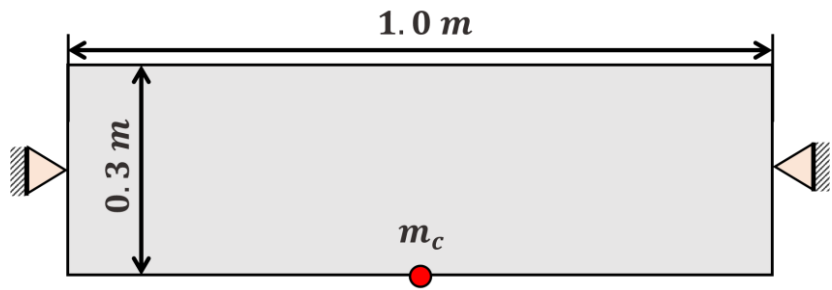

**Fig. 12** Problem setting illustration of the second example.

The optimization history for the 10-class multiscale design is shown in **Fig. 13.** The first natural frequency is a simple eigenvalue during the whole process and does not encounter any mode switch. Optimal full structures using different methods are shown in **Fig. 14**.

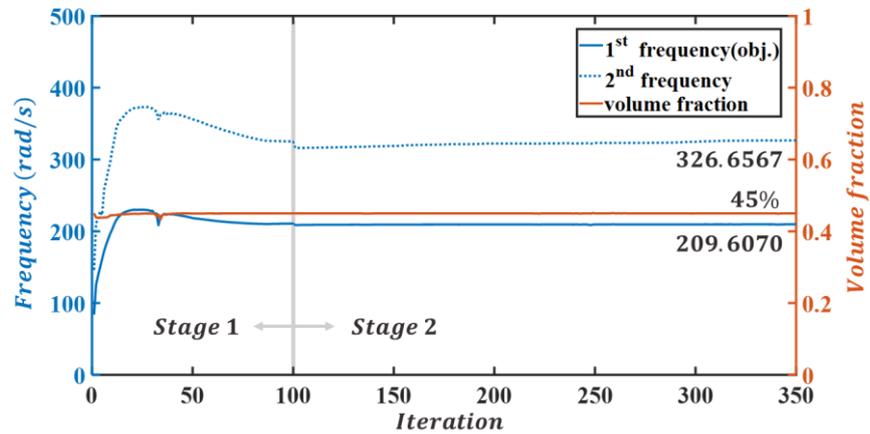

**Fig. 13** Optimization history of the 10-class design for the second example.

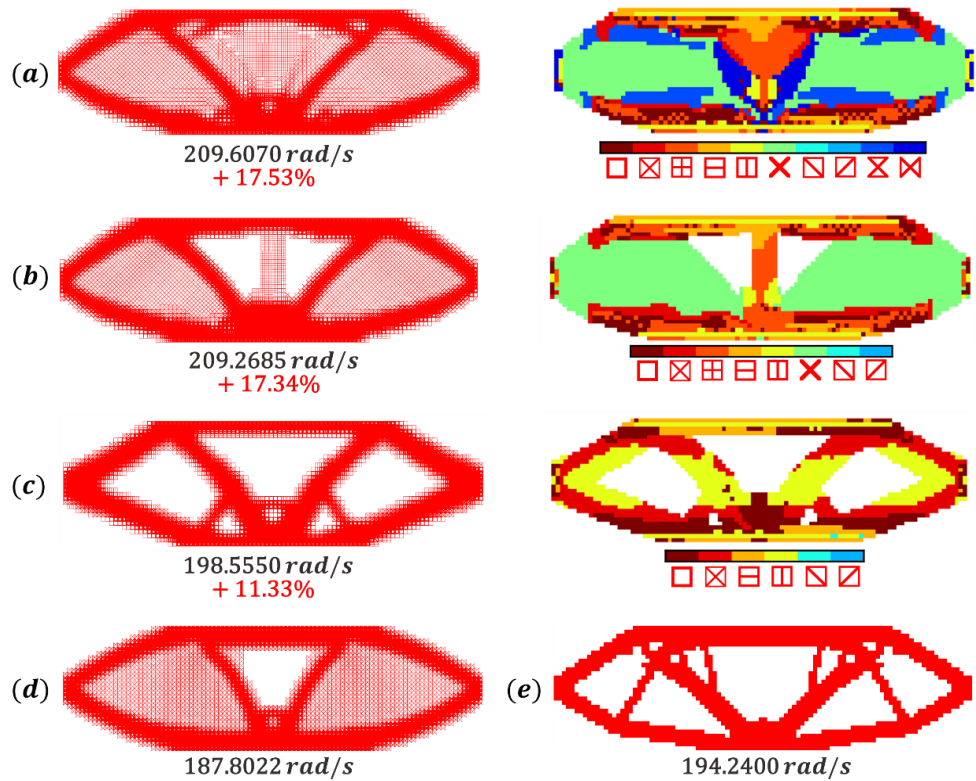

**Fig. 14** Optimal designs via different methods. The value of the first natural frequency and the improvement over single-scale design are marked under the full structure. (a) 10-class design (left) and its class distribution (right), (b) eight-class design (left) and its class distribution (right), (c) six-class design (left) and its class distribution (right), (d) single-class design, (d) single-scale design via SIMP.

Likewise, the three multiclass designs all perform better than single-scale and single-class designs. Among these structures, the one obtained by using a single class of microstructure has the lowest frequency value. As expected, the number of classes used in the optimization has a large impact on the performance and shape of the final design. The more classes we include in the optimization, the larger the gain of the multiclass design over the single-scale design. Interestingly, although the 10-class design uses a large amount of class I microstructures, the performance of the eight-class design only has a small decrease when we exclude class I from the optimization process. Regions originally filled by class I in the 10-class structure are replaced by class F in the 8-class structure. It implies that classes I and F could accommodate similar property requirements in the structure. This matches well with the fact that they are mutual nearest neighbors in the latent space, providing additional validation of the proposed LVGP-SoS model. Similarly, after excluding class F in the 6-class design, classes B and E will fill those regions originally occupied by class F so that the structure can still outperform single-class and single-scale designs.

**4.3. Cantilever Beam with Two Concentrated Masses**

In the last example, we explore the adaptivity of the proposed framework for designs with multiple concentrated masses under different boundary conditions. The design region is a cantilever beam with its left end fixed. We consider two different cases for the distribution of concentrated masses, as shown in **Fig. 15**.

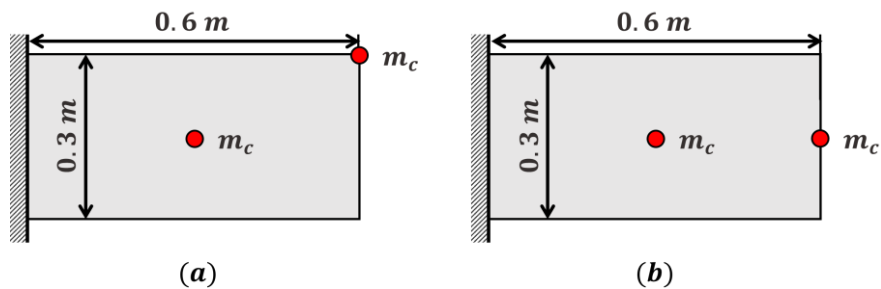

**Fig. 15** Problem setting illustration of the third example.

For the first case illustrated in **Fig. 15 (a)**, the two concentrated masses are imposed on the center and the upper-right corner. In contrast, we move one of the masses from the upper-right corner to the center of the right edge in the second example shown in **Fig. 15 (b)**. In both cases, we set each lump mass to be $m_c = 1500\ kg$. This $0.6\ m \times 0.3\ m$ beam is equally discretized into $60 \times 30$ quadrilateral elements. The volume fraction of the full structure is set to be 40%. Single-scale, single-class, and multi-class optimizations are performed for these two cases, respectively, with the result shown in **Fig. 16** and **Fig. 17**.

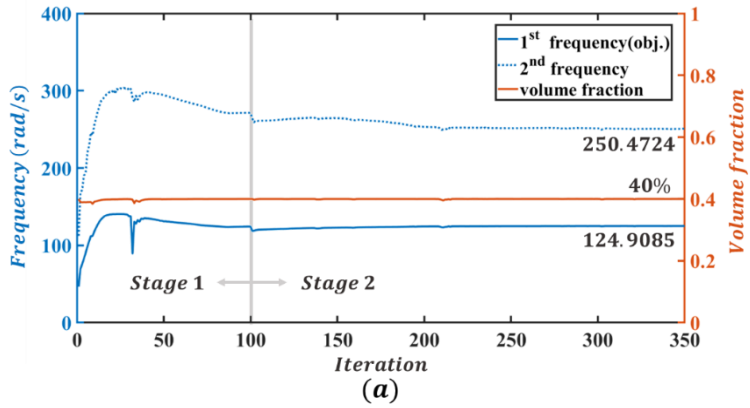

(a)

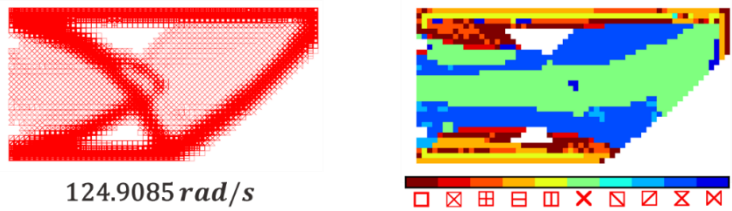

(b)

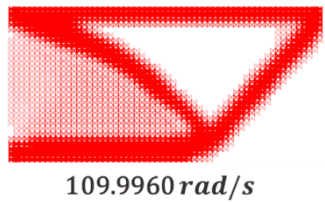 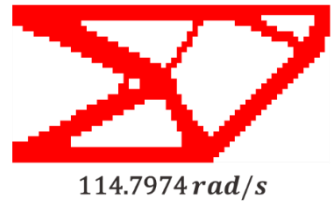

(c) (d)

**Fig. 16** Design results of the first cantilever beam, (a) is the optimization history of the multi-class design, (b) shows the assembled multi-class design (left) and its class distribution (right), (c) is the optimal single-class design, and (d) is the single-scale design via SIMP.

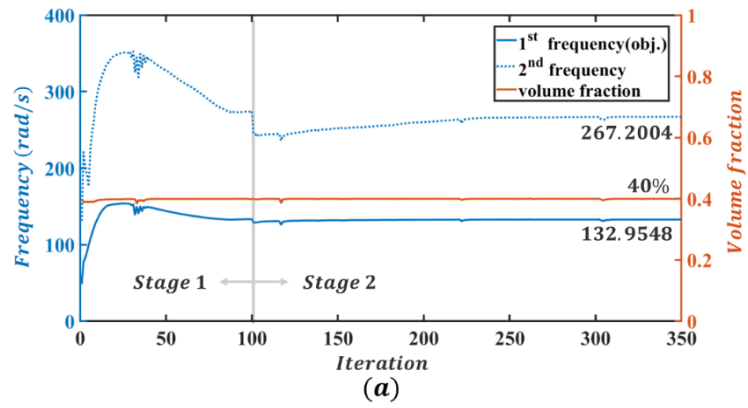

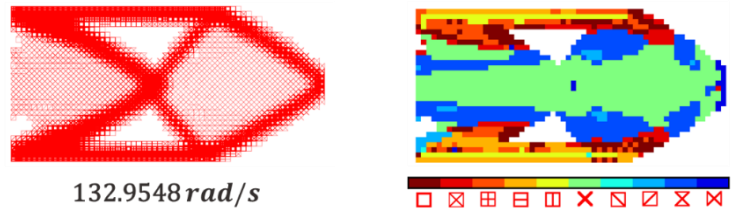

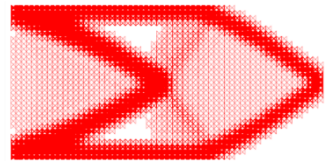
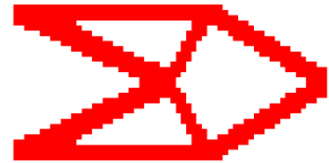

**Fig. 17** Design results of the second cantilever beam, (a) is the optimization history of the multi-class design, (b) shows the assembled multi-class design (left) and its class distribution (right), (c) is the optimal single-class design, and (d) is the single-scale design via SIMP.

From the optimization history, it can be noted that the sensitivity value remains valid during the whole process since the algebraic multiplicity of the first eigenvalue equals one during the whole process. In both cases, we can reach the same conclusion that multiclass designs have the best performance while single-class designs are among the worst. It is interesting to note that the distribution of different classes will change adaptively to accommodate different positions of the concentrated masses. This demonstrates the effectiveness and adaptivity of the proposed method.

## 5. CONCLUSIONS

We developed a data-driven multiscale TO method that can concurrently explore multiple microstructure classes and associated geometric parameters to maximize the natural frequencies of cellular structures. This method is built on a newly developed latent variable Gaussian process that can transform qualitative classes into continuous latent vectors, forming a unified quantitative design space with the associated geometric parameters. By introducing the sum of separable kernels, this new model can better accommodate the normal-shear coupling terms in the stiffness tensor, achieving a more accurate surrogate model for the material law. Based on this LVGP-SoS model, a new data-driven multiscale TO method with multiclass microstructures is devised by including two-dimensional latent variables as extra design variables in the classical density-based TO method.

The proposed method has been applied to several fundamental frequency maximization problems with various boundary conditions and lump mass distribution cases. From the design examples, we demonstrate that data-driven designs with a single class of microstructure do not provide any benefit over single-scale design in terms of frequency maximization. In contrast, multi-class designs obtained with the proposed method always have better performances than both single-scale and single-class designs. The more classes of microstructures considered in the optimization, the greater the advantage will be. Moreover, the distribution of different classes in the full structure can adaptively change to accommodate different local requirements under different boundary conditions. This underlines the importance of including multiple microstructure design concepts in the pre-computed library for better performance and adaptivity. Meanwhile, due to the use of reduced-dimensional latent variable representation and an inexpensive surrogate model, our method can have an efficiency comparable to the TO that only considers macroscale design.

To further our research, we plan to extend the proposed method to other multiscale design applications, such as thermal compliance minimization and dynamic response optimization. Even though we only include the class of microstructures as the only qualitative variable in this work, the same framework can be easily extended to consider more qualitative variables, such as the type of base materials.

## ACKNOWLEDGMENTS

We are grateful for support from the NSF CSSI program (Grant No. OAC 1835782) and Center for Hierarchical Materials Design (ChiMaD NIST 70NANB19H005). Liwei Wang acknowledges support from the Zhiyuan Honors Program for Graduate Students of Shanghai Jiao Tong University for his predoctoral visiting study at Northwestern University. Yu-Chin Chan thanks the NSF Graduate Research Fellowship (Grant No. DGE-1842165).

## DATA AVAILABILITY

The raw and processed data required to reproduce these findings are available to download from https://ideal.mech.northwestern.edu/research/software/.